\documentclass[seceqn,secthm,final]{elsart}
\journal{}

\usepackage{amssymb,amsmath}
\usepackage{graphicx}

\renewcommand{\theenumi}{\roman{enumi})}

\begin{document}

\begin{frontmatter}
\title{On the streamlines and particle paths of gravitational water waves}
\author{Mats Ehrnstr\"om}
\address{Department of Mathematics, Lund University, PO Box 118, 
221 00 Lund, Sweden.}
\ead{mats.ehrnstrom@math.lu.se}

\begin{abstract}
We investigate steady symmetric gravity water waves on finite depth. For non-positive vorticity it is shown that the particles display a mean forward drift, and for a class of waves we prove that the size of this drift is strictly increasing from bottom to surface. We also provide detailed information concerning the streamlines and the particle trajectories. This includes the case of particles within irrotational waves.
\end{abstract}


\end{frontmatter}

\section{Introduction} \label{sec:intro}

This paper is concerned with the streamlines and the particle trajectories of steady gravity water waves on finite depth. Such water waves are one of the most common wave formations at sea. As a result of dispersion wind-generated gravity waves eventually sort themselves out \cite{MR1629555,MR642980}. Larger waves move faster than smaller ones and swell is generated: approximately two-dimensional wave-trains of periodic and symmetric waves moving with constant speed across the sea. The exact mathematical theory for such waves is well established, in particular for irrotational currents \cite{MR2097656,MR1422004}. Those model very well the situation when the waves propagate into a region of still water. There are however experimental evidence that for some situations such a model is inadequate \cite{SCJ01}. One example is tidal flow, which is more correctly modelled by waves entering a rotational current of constant vorticity \cite{MR985439}. Therefore the importance of water motion with a non-vanishing curl --- i.e. in the presence of vorticity --- has recently come to draw a lot of attention (see e.g. \cite{MR2264220,E06a,Hur2007,wahlen:921}). For us it is of relevance that for arbitrary vorticity distributions there exist symmetric waves \cite{MR2027299}, and that any wave for which the surface profile is monotone from crest to trough necessarily is symmetric \cite{CEW06}.

In a number of recent articles the exact behaviour of the fluid particles within such waves have been investigated \cite{MR2257390,CEV06,CV06,23405,MR2287829}. The background is the following.
For over a century it has been known that the very first approximation of steady irrotational gravity water waves display closed elliptic particle trajectories \cite{Stokes49}. However, as was first noted in \cite{CV06}, a thorough study of the linearized system shows that the particle paths indeed have an elliptic shape, but are not closed. This is so for other types of waves too, as has been shown in \cite{CEV06,MR2287829}. Since the linearized problem can be solved explicitly, details of the particle paths can be more easily studied. In particular it can be seen that all particles traverse elliptic orbits.  When vorticity is present things are not as transparent, not even for linear waves on a current of constant vorticity. In \cite{Ehrnstrom2007b} it is e.g. shown that when the size of the vorticity is large, the particle paths of linear waves need not all be circular; some particles may move constantly forward along with the wave. Near the flat bed the particles however always behave like the classical first approximation: they move slightly forward in circular shapes. As will be discussed below this is all in relation to some reference speed, i.e. the generalized Stokes requirement \eqref{eq:myStokes}.

For exact water waves the details are far more elusive since closed expressions are not available. The investigations \cite{MR2257390,23405} 
show that even for exact irrotational water waves the particles display a mean forward drift. They also assert that for irrotational Stokes waves on finite as well as infinite depth all particles above the bed move in circular orbits. In this paper we show that the mean forward drift is preserved for all negative vorticity distributions. For irrotational waves and small enough rotational waves, we are able to show that this forward drift is strictly increasing from bottom to trough. A proof of this for linear waves without vorticity was given in \cite{CEV06}. In addition we establish some very nice properties of the velocity field and the particle paths, in particular so for irrotiational waves. We have not been able to confirm the circular orbital shapes for rotational waves. 

The novelty of our approach lies in the fact that via maximum principles we establish precise pointwise information about the velocity field and its derivatives within the entire fluid domain. In this way we extend the propositions in \cite{MR2257390,23405}, providing further understanding even for the irrotational case. For steady rotational waves, to our knowledge this is the first investigation of its kind apart from \cite{Ehrnstrom2007b}. The paper is organized as follows. Section~\ref{sec:formulation} gives the mathematical background, while the main results are proved in Sections~\ref{sec:prop} and~\ref{sec:forward}. A synthesis and an analysis of the particle paths are given in Section~\ref{sec:trajectories}, presented as two, hopefully illustrative, examples. At the end Section~\ref{sec:comments} contains some short comments on the results.

\section{Mathematical formulation}\label{sec:formulation} 
Let $d > 0$ be the depth below the mean water level $y=0$, so that the flat bottom can be described by $y = -d$. 
The \emph{free surface} can then be represented by a function 
\begin{equation*}
\eta \in C^3(\Rset,\Rset).
\end{equation*}
We require that $\eta(0) = \max_{x\in\Rset} \{\eta(x)\}$ be the vertical coordinate of the crest (unique within a period). Naturally $\min_{x\in\Rset} \{\eta(x)\} > -d$ so that the trough is above the flat bed $y = -d$. In this paper we shall be concerned with the nontrivial case when $\max \eta > \min \eta$. The wave is steady of period $L>0$ --- without loss of generality we may take $L = 2\pi$ ---  and we require the surface profile to be monotone between crests and troughs. It is therefore symmetric around the crest \cite{CEW06}, and we have that 
\begin{equation*}
\eta(x+2\pi) = \eta(x), \quad \eta(x) = \eta(-x), \quad\text{ and}\quad \eta^\prime(x) < 0 \text{ for } x \in (0,\pi).
\end{equation*}
We let $\Omega_\eta$ denote the \emph{fluid domain} and define it as the interior of its boundary
\begin{equation*}
\partial \Omega_\eta \equiv \{y=-d\} \cup \{x,\eta(x))\}_{x\in\Rset}.
\end{equation*}
A solution to the water wave problem is then definied as a function $\psi \in C^2(\overline\Omega_\eta)$ such that
\begin{align}\label{eq:problem} 
\begin{cases}
\Delta \psi = -\gamma(\psi), &(x,y)\in\Omega_\eta\\
|\nabla\psi|^2 + 2gy = C,\qquad & y=\eta(x)\\
\psi = 0, & y=\eta(x)\\
\psi = -p_0, &y = -d,
\end{cases}
\end{align}
that is even and $2\pi$-periodic in the $x$-variable. In \eqref{eq:problem} $p_0$ is called the \emph{relative mass flux}, the \emph{vorticity function} $\gamma\colon [0,-p_0] \rightarrow \Rset$ is continuously differentiable, $g > 0$ is the gravitational constant, and $C$ is a constant related to the energy. The setting is that of gravitational water waves, meaning that the influence of capillarity is neglected in~\eqref{eq:problem}, and the water is assumed to be inviscid. The \emph{stream function} $\psi$ is defined (up to a constant) by
\begin{equation}\label{eq:psi_y}
\psi_x = -v, \qquad \psi_y = u-c < 0,
\end{equation}
where $u,v$ are the horizontal and the vertical velocity, respectively, and $c > 0$ is the constant horizontal speed of propagation. The notion of relative mass flux introduced in \cite{MR2027299} captures the physical fact that the amount of water passing any vertical line is constant troughout the fluid domain:
\begin{equation*}
\int_{-d}^{\eta(x)} \left(u(x,y)-c\right)\,dy = p_0, \qquad x \in \Rset,
\end{equation*}
holds since $u-c = \psi_y$, and $\psi$ is constant on the surface $y=\eta(x)$ as well as on the bottom $y = -d$.

Provided that $u - c = \psi_y < 0$, the system (\ref{eq:problem}) can be deduced from the Euler equations (see e.g. \cite{MR2027299,MR2130838} for a more detailed discussion). This assumption is supported by physical measurements \cite{MR642980}: for a wave not near breaking or spilling, the speed of an individual fluid particle is far less than that of the wave itself. For irrotational waves it is however known that there exist so called highest waves for which the crest is a stagnation point, i.e. $\nabla \psi = 0$. While the exact problem is still open for waves with vorticity  \cite{Varvaruca2007a}, there are indications that for some classes of vorticity there do exist steady waves with particle layers not satisfying $\psi_y < 0$ \cite{Ehrnstrom2007b}.  In this paper we shall however consider only waves that are not near breaking or stagnation, so that $\psi_{y} < 0$ in $\overline\Omega_\eta$.

A hodograph transform converts the free boundary problem (9) into a 
problem with a fixed boundary. Let us express the height 
\[
h \equiv y+d
\]
above the flat bed in terms of the new space variables
\begin{equation}\label{eq:qp}
q \equiv x,\qquad p \equiv -\psi.
\end{equation}
Notice that $\psi_y<0$ so that \eqref{eq:qp} is a local change of variables, with 
\[
h_q \equiv -\frac{\psi_x}{\psi_y}=\frac{v}{u-c},\qquad h_p \equiv - \frac{1}{\psi_y}=\frac{1}{c-u}.
\] 
The above local coordinate transform is actually a global change of variables (see \cite{MR2027299}) 
so that we can transform the problem \eqref{eq:problem} to these variables to obtain 
\begin{equation} \label{eq:hproblem}
\begin{cases}
(1+h_q^2)\,h_{pp}-2h_ph_qh_{pq}+h_p^2h_{qq}+\gamma(-p)\,h_p^3=0
\quad\hbox{in}\quad p_0<p<0,\\
1+h_q^2 + (2gh-Q)\,h_p^2=0\quad\hbox{on}\quad p=0,\\
h=0\quad\hbox{on}\quad p=p_0,
\end{cases}
\end{equation}
with $h$ even and of period $2\pi$ in the $q$ variable. This is an elliptic equation (since $h_p>0$) with a nonlinear boundary condition.  Instead of studying \eqref{eq:problem} in the domain $\Omega_\eta$, which depends on $\eta$, we investigate \eqref{eq:hproblem} in the fixed rectangle $R \equiv (-\pi,\pi) \times (p_0,0)$, looking for functions $h \in C^2(\overline{R})$ that are $2\pi$-periodic in $q$. Notice that knowing $h(q,0)$ is equivalent to knowing the free surface $y = \eta(x)$ as $h(q,0) = \eta(q)+d$. 

Let $(x,\sigma(x))$ denote the parametrization of a (general) streamline 
\[
\{(x,y)\colon \psi(x,y) = -p\}.
\]
Note that since $\psi_{y} < 0$ the above parametrization is sensible, and we have 
\[
\sigma^\prime(x) = - \frac{\psi_{x}(x,\sigma(x))}{\psi_{y}(x,\sigma(x))} = h_{q}(q,p).
\]

To normalize the reference frame Stokes made a now commonly accepted proposal. In the case of irrotational flow he required that the horizontal velocity should have a vanishing mean over a period. Stokes' definition of the wave speed unfortunately cannot be directly translated to waves with vorticity, a consequence of that $\mathrm{div}\, \nabla \psi \not\equiv 0$, so that $\psi_y$ has different means at different depths in view of the Divergence theorem. In the setting of periodic waves with vorticity we propose the requirement
\begin{equation}\label{eq:myStokes}
\int_{-\pi}^\pi u(x,-d)\,dx = 0,
\end{equation}
a ``Stokes' condition'' at the bottom. This is consistent with deep-water waves (cf. \cite{Ehrnstrom2007d}), and is also the choice made in \cite{MR985439}. Calculations performed on linear water waves with constant vorticity indicate that this is the natural choice, since that and only that choice recovers the well established bound $\sqrt{gh}$ for the wave speed \cite{Ehrnstrom2007b}. We emphazise that \eqref{eq:myStokes} is only a convention for fixing the reference frame; except from the assertion of forward drift it does not change anything in this paper. Without such a reference it is however meaningless to discuss whether physical particle paths are closed or not.

\section{Streamlines and the horizontal velocity}\label{sec:prop}
In this section three propositions are given. The proofs are collected at the end of the section. Lemma~\ref{lemma:prop} mainly contains content known within the field, which we make use of in the rest of the paper. Lemma~\ref{lemma:horizontal} gives detailed information about the horizontal velocity throughout the fluid domain. Its Corollary~\ref{cor:eta} gives a lower bound for the surface of a large class of gravity waves. 

\begin{lem}\label{lemma:prop}\
\begin{enumerate}
\item\label{lemma:hq} Every streamline satisfies $\sigma^\prime < 0$ for $x \in (0,\pi)$, and the maximal steepness of the streamlines is a strictly monotone function of depth. 
\item\label{lemma:hp} For $\gamma^\prime \geq 0$ and $\gamma \leq 0$, the maximal horizontal velocity, $\max\limits_{x\in \Rset} u(x,\sigma(x))$, is strictly increasing from bottom to surface.
\item\label{lemma:v} The vertical velocity is strictly positive for $x \in (0,\pi)$, and if $\gamma^\prime \leq 0$, then $\max\limits_{x\in \Rset} |v(x,\sigma(x))|$ is strictly increasing from bottom to surface.
\end{enumerate}
\end{lem}

The fact that $h_{q} = \sigma^\prime$ has a strict sign was first noticed in \cite{MR2027299}. The second proposition is a slightly different version of the simple fact that $\Delta\psi_y = -\gamma^\prime(\psi)\psi_y \geq 0$. The proof here presented is new. The requirement on $\gamma^\prime$ in \ref{lemma:v} dates back to \cite{MR2256915}. Our proof techniques rely heavily on sharp maximum principles. We refer the reader to the excellent sources \cite{MR1751289,1109.35022}.

The following lemma gives detailed information about the horizontal velocity.

\begin{lem}[The horizontal velocity]\label{lemma:horizontal}\
\begin{enumerate}
\item\label{lemma:eugen} If $\gamma \leq 0$, then the horizontal velocity $u$ is non-increasing from crest to trough, i.e. 
\begin{equation}\label{eq:Dxu}
\mathrm{D}_x u(x,\eta(x)) \leq 0 \quad\text{ for }\quad x \in (0,\pi).
\end{equation}
\item\label{lemma:hqp}
If $\gamma = 0$ and $|\eta^\prime| \leq 1/\sqrt{3}$, then along any streamline $(x,\sigma(x))$ holds
\[
\mathrm{D}_x u(x,\sigma(x)) < 0 \quad\text{ for }\quad x \in (0,\pi),
\]
and the pointwise steepness of the streamlines is everywhere decreasing with depth.
\item\label{lemma:localpsixy} 
If $\gamma(0) \geq 0$ and $\gamma^\prime, \gamma^{\prime\prime} \leq 0$ then 
\[
\partial_x u(x,y) < 0 \quad\text{ in }\quad (0,\pi) 
\]
for waves in a neighbourhood of the bifurcation point in \cite{Constantin2007}. 
\item\label{lemma:u}
If the horizontal velocity attains its maximum at the surface, then it does so either at the crest, or at the concave part of the surface where 
\[
(c-u) \gamma < g = -\eta^{\prime\prime}(c-u)^2.
\]
\end{enumerate}
\end{lem}

Part \ref{lemma:eugen} was proved in \cite{Varvaruca2007}, building on earlier ideas for the irrotational \cite{MR1422004} and the rotational \cite{Constantin2007} cases. 

Concerning part \ref{lemma:hqp} it is rather wonderful that the bound $|h_{q}| \leq 1/\sqrt{3}$ appears in the proof of this assertion. It is the natural number since it describes the interior angle at the crest of Stokes highest wave, see \cite{Varvaruca2007a}. Once we proved Lemma~\ref{lemma:prop}~\ref{lemma:hqp}, the equality \eqref{eq:hdiffeomorphism} implies that $\mathrm{Det} \, \mathcal{J} (\nabla h) \neq 0$, so that the mapping $(q,p) \mapsto (h_{q},h_{p})$ is everywhere locally a $C^1$-diffeomorphism. The rotational property however seems to destroy that feature. 

Part \ref{lemma:localpsixy} equivalently states that $\partial_y v  < 0$ within the half-period $(0,\pi)$, so that the vertical velocity behaves similarly to its linear approximation for constant vorticity, $v(x,y) \sim \sin(x)\sinh(y)$ (cf. \cite{Ehrnstrom2007b}). 

Part \ref{lemma:u} limits the possible classes of waves with the maximal horizontal velocity bounded away from the crest. Note also that the equality of this claim is based solely on the surface conditions of \ref{eq:problem}, and hence is unrelated to e.g. periodicity, symmetry, and depth. 

\begin{cor}[The surface]
\item\label{cor:eta}
If \eqref{eq:Dxu} holds, then for $x \in (0,\pi)$  we have the uniform bound $\eta^{\prime\prime} \geq -\frac{g}{C-2g\eta(0)}$, together with
\begin{equation*}
\eta^\prime(x) \geq -\frac{gx}{C-2g\eta(0)}, \quad\text{ and }\quad \eta(x) \geq \eta(0) - \frac{g x^2}{2(C-2g\eta(0))}.
\end{equation*}
\end{cor}

It should be pointed out that Corollary~\ref{cor:eta} is based solely on the surface conditions, hence valid for all types of gravity waves.

\begin{pf*}{Proof of Lemma~\ref{lemma:prop}}
\begin{enumerate}
\item For details see \cite[Eq. (5.18)]{MR2027299}. The key idea is that $h_{q}$ is annihilated by 
\begin{equation}\label{eq:hqequality}
\left(1+h_q^2\right)\partial_p^2 - 2 h_p h_q \partial_p \partial_q + h_p^2 \partial_q^2  + 2h_{q} h_{pp} \partial_q  + \left[3\gamma(-p)h_{p}^2-2h_{q} h_{pq}\right] \partial_p.
\end{equation}
The second statement follows from applying the strong maximum principle to subdomains $(0,\pi) \times (-d,p)$ of this half-period.
\item Consider $h_p = 1/(c-u) > 0$. Since $h_p$ belongs to the kernel of the uniformly elliptic operator
\begin{equation}\label{eq:hpequality}
\left(1+h_{q}^2\right)\partial_{pp} - 2h_{q} h_{p} \partial_{qp} + h_{p}^2\partial_{qq} -2h_{p} h_{qp} \partial_{q} + h_{p}\left(2h_{qq}+3\gamma h_{p}\right)\partial_{p} - \gamma^\prime h_{p}^2,
\end{equation}
the strong maximum principle implies that $\max u$ is never attained in the interior of any $C^2$-subdomain of the fluid. Since 
\[
u_{y} = \psi_{yy} = -\gamma(-p_0) \geq 0
\] 
at the flat bed, it is a consequence of the Hopf boundary point lemma that $u$ does not attain its maximum on the bottom. The proposition then follows by periodicity.
\item Since $\sigma^\prime = -\psi_x/\psi_y$ and $\psi_y < 0$, the positivity of $v$ is a direct consequence \ref{lemma:hq}. Note also that $v$ vanishes on the flat bed. Since $v \in \mathrm{Ker}\{\Delta + \gamma^{\prime}(\psi)\}$ we may apply the strong maximum principle to any subdomain $(0,\pi)\times(-d,\sigma(x))$. Symmetry yields the assertion.
\end{enumerate}
\end{pf*}

\begin{pf*}{Proof of Lemma~\ref{lemma:horizontal}}\
\begin{enumerate}
\item See \cite{Varvaruca2007}[Thm 2.2]
\item We need to show that $h_{qp} < 0$ everywhere in $q \in (0,\pi)$. Recall that $h_{p} = 1/(c-u)$ and that each value of $p$ corresponds to a streamline. Therefore Lemma~\ref{lemma:prop}~\ref{lemma:eugen} shows that $h_{qp} \leq 0$ along the surface for $q \in [0,\pi]$. At the bottom we have $h_{qp} < 0$ for $q \in (0,\pi)$. This follows from Lemma~\ref{lemma:hq} and the Hopf boundary point lemma. Along the vertical sides where $q = 0$ and $q = \pi$, symmetry implies that $h_{qp} = 0$. We shall  establish that the strong maximum principle holds for $h_{qp}$ in $q \in [0,\pi]$, according to which $h_{qp} \geq 0$ at an interior point of the half-period would force $h_{qp} = const$ everywhere. 
  
We begin by differentiating \eqref{eq:hqequality} with respect to $p$. In that expression appears expressions involving $h_{qqq}$ and  $h_{ppp}$. Those are substituted using \eqref{eq:hqequality} and \eqref{eq:hpequality}, so that on the level of third and fourth order derivatives of $h$, only such that are derivatives of $h_{qp}$ appears. Making repeated use of the elliptic equality in \eqref{eq:hproblem} it is possible to write the obtained expression as the following identity:
\begin{multline}\label{eq:hqp}
\Bigg[\left(1+h_q^2\right)\partial_p^2 - 2 h_p h_q \partial_p \partial_q + h_p^2 \partial_q^2 + \left( 4h_q h_{pp} - 2 h_p h_{qp} - \frac{2 h_q h_p^2 h_{qq}}{1+h_q^2}\right)\partial_q\\
+ \left(3\gamma h_p^2 - 2 h_q h_{qp} + \frac{4 h_q^2 h_{qq} h_p}{1+h_q^2} - \frac{2 h_{pp} (1+h_q^2)}{h_p} \right)\partial_p \\
+ \frac{2(h_{qq} h_{pp} - h_{qp}^2)(1-3h_q^2) - 3\gamma^\prime h_p^2}{1+h_q^2}\Bigg]  h_{qp}\\
= \frac{2 h_q h_p^2}{1+h_q^2}\left( \gamma (h_{qq} h_{pp} + 2 h_{qp}^2)-\gamma^\prime h_p h_{qq}\right).
\end{multline} 
By multiplying the elliptic equation in \eqref{eq:hproblem} with $h_{pp}$ and then completing the squares, we obtain for $\gamma = 0$ that
\begin{equation}\label{eq:hdiffeomorphism}
0 = (h_{q} h_{pp} - h_{p} h_{qp})^2 + h_{pp}^2 + h_{p}^2(h_{qq}h_{pp}-h_{qp}^2),
\end{equation}
which forces $h_{qp}^2 \geq h_{qq}h_{pp}$ everywhere in the fluid. Finally, Lemma~\ref{lemma:prop}~\ref{lemma:hq} and the assumption guarantees that $h_{q}^2 \leq 1/3$ everywhere. 
\item
We shall determine $\psi_{xy}$ in a manner similar to that of  \cite{Ehrnstrom2007c}. Since $\psi = 0$ along the surface, we have $\psi_x = -\dot\eta \psi_y$, and insertion into the Bernoulli surface condition of \eqref{eq:problem} yields that
\begin{equation}\label{eq:psiy2}
\psi_y^2 = \frac{C-2g\eta}{1+\dot\eta^2}.
\end{equation}
It follows that  
\begin{equation*}
\psi_{xy} + \dot\eta \psi_{yy} = - \partial_x \sqrt{(C-2g\eta)/(1+\dot\eta^2)}.
\end{equation*}
On the other hand, we may differentiate $\psi_x = \dot\eta \psi_y$ once more, obtaining
\begin{equation*}
\psi_{xx} + \dot\eta^2 \psi_{yy} + 2\dot\eta \psi_{xy} + \ddot\eta \psi_y  = 0.
\end{equation*}
We now know the first derivatives of $\psi$ in terms of $\eta$ and in order to determine the three second derivatives we only need one more equality. This is supplied by
\begin{equation*}
\psi_{xx}+\psi_{yy} = -\gamma(0).
\end{equation*}
The calculation, which is tedious but easily carried out by hand, yields that at the surface $\psi_{xy}$ can be determined as
\begin{equation}\label{eq:psixy}
\psi_{xy} = \frac{\dot\eta \left(2\ddot\eta(C-2g\eta) + \left(1-\dot\eta^4\right)g + \gamma(0) \sqrt{C-2g\eta}\left( 1+\dot\eta^2\right)^{3/2} \right)}{\left( 1+\dot\eta^2\right)^{5/2}\sqrt{C-2g\eta}}.
\end{equation}
Since $0 \leq C - 2g\eta \leq C + 2gd$ we see that for small enough $\eta^\prime, \eta^{\prime\prime}$ the sign is determined by $g$ (remember that $\gamma(0) \geq 0$ by assumption), so that $\psi_{xy} \leq 0$ at the surface. But $\psi_{xy}$ obeys the maximum principle, 
\[
\left(\Delta + \gamma^\prime\right)\psi_{xy} = -\gamma^{\prime\prime}\psi_x \psi_y \geq 0,
\]
so that a nonnegative maximum in $0\leq x \leq \pi$ is attained either on the flat bed or at the surface (since $\psi_{xy} = 0$ for $x = k\pi$, $k \in \Zset$). Since Lemma~\ref{lemma:prop}~\ref{lemma:v} and the Hopf boundary point lemma forces that $\psi_{xy} < 0$ for $y = -d$, the proof is complete. 
\item 
In the spirit of \cite{Ehrnstrom2007c}, and according to \eqref{eq:psiy2}, differentiation along the surface gives
\begin{equation}\label{eq:Dxpsiy2}
D_x \psi_y^2(x,\eta(x)) = \frac{2\dot\eta \left[(2g\eta-C)\ddot\eta -g\left(1+\dot\eta^2\right) \right]}{\left(1+\dot\eta^2\right)^2}.
\end{equation}
Thus a maximum implies that either $\dot\eta = 0$ or ${(2g\eta-C)}\ddot\eta = g\left(1+\dot\eta^2\right)$. With a little manipulation, half of the result is obtained by substituting the second expression into the Bernoulli surface condition of \eqref{eq:problem}. To obtain the inequality, first note that if the maximum is attained at the surface, then $\psi_{yy} > 0$ at that point. But (cf. \cite{Ehrnstrom2007c})
\begin{equation}\label{eq:psiyy}
\psi_{yy} = \frac{\ddot\eta(C-2g\eta)\left(\dot\eta^2-1\right) + 2g\dot\eta^2\left(1+\dot\eta^2\right)  - \gamma(0) \sqrt{C-2g\eta}\left( 1+\dot\eta^2\right)^{3/2}}{\left( 1+\dot\eta^2\right)^{5/2}\sqrt{C-2g\eta}},
\end{equation}
so we need only substitute ${(2g\eta-C)}\ddot\eta = g\left(1+\dot\eta^2\right)$ into that expression to get 
\begin{equation}\label{eq:eval2}
\psi_{yy}\big|_{\ddot\eta = \frac{g\left(1+\dot\eta^2\right)}{(2g\eta-C)}} = \frac{g\sqrt{1+\dot\eta^2} -\gamma(0) \sqrt{C-2g\eta} }{\left( 1+\dot\eta^2\right)\sqrt{C-2g\eta}} \geq 0.
\end{equation}
\end{enumerate}
\end{pf*}

\begin{pf*}{Proof of Corollary~\ref{cor:eta}}
Recall \eqref{eq:Dxpsiy2}. It follows from the assumption that 
\begin{equation}\label{eq:logeq}
\frac{2\eta^\prime \eta^{\prime\prime}}{1+{\eta^\prime}^2} \leq \frac{-2g\eta^\prime}{C-2g\eta}, \quad\text{ meaning }\quad \frac{d}{dx}\log\left(\frac{1+{\eta^\prime}^2}{C-2g\eta}\right) \leq 0.
\end{equation}
This can be integrated to
\begin{equation}\label{eq:firstint}
\frac{1+{\eta^{\prime}}^2}{C-2g\eta} \leq \frac {1}{C-2g\eta(0)}.
\end{equation}
Since $\eta^\prime < 0$ in $(0,\pi)$ we may rearrange to obtain
\begin{equation}\label{eq:rearranged}
\frac{-\eta^{\prime}}{\sqrt{2g(\eta(0)-\eta(x))}} \leq \frac{1}{\sqrt{C-2g\eta(0)}}.
\end{equation}
Now the assertion concerning $\eta$ is established by integrating \eqref{eq:rearranged}. The bound on $\eta^\prime$ then follows from employing the lower bound on $\eta$ to \eqref{eq:firstint}. Finally, the uniform bound on $\eta^{\prime\prime}$ is immediate from combining the left-hand side of \eqref{eq:logeq} with \eqref{eq:firstint}. 
\end{pf*}

\section{The forward drift}\label{sec:forward}
In this section we prove that for waves of non-positive vorticity the forward drift of the fluid particles is everywhere positive within the fluid domain. For irrotational waves and small enough rotational waves, it is shown that the drift is monotone from bottom to surface. The proofs are collected at the end of the section.

\begin{thm}\label{thm:forward}\
\begin{enumerate}
\item
For $\gamma \leq 0$ there are no closed particle trajectories above the flat bed. In particular all fluid particles display a mean forward drift.
\item
If $\gamma = 0$ with $|\eta^\prime| \leq 1/\sqrt{3}$ then the mean forward drift is strictly increasing from bed to surface. 
\item
If $\gamma < 0$ then for all waves in a neighbourhood of the bifurcation point found in \cite{MR2027299}, the mean forward drift is strictly increasing from bed to surface. 
\end{enumerate}
\end{thm}

\begin{lem}\label{lemma:bound}
For $(x,\sigma(x))$ a non-trivial streamline, and $\gamma \leq 0$, the quantity 
\[
\int_0^\pi |\psi_y(x,\sigma(x))| \left(1+{\sigma^\prime}^2(x)\right) \,dx 
\]
is non-decreasing with depth, and
\begin{equation}\label{eq:cpi}
\int_0^\pi |\psi_y(x,\sigma(x))|\,dx < c\pi.
\end{equation}
\end{lem}

\begin{pf*}{Proof of Lemma~\ref{lemma:bound}}
For two streamlines $(x,\sigma_1(x))$ and $(x,\sigma_2(x))$ with $\sigma_1(x) < \sigma_2(x)$, let
\[
\Sigma \equiv (0,\pi) \times (\sigma_1(x),\sigma_2(x)).
\] 

According to the Divergence theorem we have
\begin{align*}
-\int_\Sigma \gamma\, dA =  \int_\Sigma \nabla \cdot \nabla \psi\, dA = \int_{\sigma_1}  \nabla \psi \cdot \frac{\nabla \psi}{|\nabla \psi |} \, ds - \int_{\sigma_2}  \nabla \psi \cdot \frac{\nabla \psi}{|\nabla \psi|} \, ds \\
= \int_0^\pi \left(\left|\nabla \psi (x,\sigma_1(x))\right|   \sqrt{1+{\sigma_1^\prime}^2(x)} -  \left|\nabla \psi (x,\sigma_2(x))\right|  \sqrt{1+{\sigma_2^\prime}^2(x)}\right)\,dx\\ 
= \int_0^\pi   \left( \left|\psi_y(x,\sigma_1(x))\right|  \left(1+{\sigma_1^\prime}^2(x)\right) -\left|\psi_y(x,\sigma_2(x))\right|  \left(1+{\sigma_2^\prime}^2(x)\right) \right)\,dx.
\end{align*}
This implies that for any nontrivial streamline $(x,\sigma(x))$, 
\[
\int_0^\pi |\psi_y(x,\sigma(x))| \left(1+{\sigma^\prime}^2(x)\right) \,dx = c\pi + \int_0^\pi \int_{-d}^{\sigma(x)} \gamma\, dA, 
\]
in view of that $\int_0^\pi \psi_y(x,-d)\,dx = -c\pi$ by the normalization \eqref{eq:myStokes}. The lemma follows. 
\end{pf*}

\begin{pf*}{Proof of Theorem~\ref{thm:forward}}\
\begin{enumerate}
\item In the physical variables, $(\mathbb{X}(t),\mathbb{Y}(t))$, a closed physical trajectory implies 
\[
(\mathbb{X}(nT),\mathbb{Y}(nT)) = (\mathbb{X}(0),\mathbb{Y}(0)), \qquad (\mathbb{\dot X}(nT), \mathbb{ \dot Y}(nT)) =   (\mathbb{\dot X}(0), \mathbb{\dot Y}(0)) 
\]
for some $T>0$ and all $n \in \Zset^+$. In the steady variables, this means
\[
x(0) - x(nT) = ncT, \qquad y(nT) = y(0).
\]
We recall that $(\mathbb{\dot X}, \mathbb{\dot Y}) = (u,v)$, so that by periodicity -- and since $\dot\sigma = 0$ and $v=0$ only for $x=n\pi$ -- this implies
\begin{equation}\label{eq:cT}
cT = 2\pi n, \quad\text{ for some }\quad n \in \Zset^+. 
\end{equation}
By periodicity $\tau = T/n$ is the time it takes a trajectorty $x(t)$ in the steady variables to pass from $x = \pi$ to $x=-\pi$. From \eqref{eq:cT} we infer that a physical particle trajectory is closed if and only if $\tau = 2\pi/c$. We also see from this reasoning that  if $\tau > 2\pi/c$, then the physical difference $\mathbb{X}(T) - \mathbb{X}(0) > 0$ when $\mathbb{Y}(T) = \mathbb{Y}(0)$, so that the particle displays a mean forward drift, and contrariwise. In our case 
\begin{multline*}
\tau/2 = t(0)-t(\pi) = -\int_0^\pi \frac{dt}{dx}\, dx\\ 
= - \int_0^\pi \frac{dx}{\dot x (x,\sigma(x))}
= \int_0^\pi \frac{dx}{c-u(x,\sigma(x))} = \int_0^\pi \frac{dx}{|\psi_y(x,\sigma(x))|}.
\end{multline*}
According to H{\"o}lders inequality
\[
\pi^2 = \left(\int_0^\pi dx\right)^2 \leq  \int_0^\pi |\psi_y(x,\sigma(x))|\,dx \,  \int_0^\pi \frac{dx}{|\psi_y(x,\sigma(x))|} < c\pi  \int_0^\pi \frac{dx}{|\psi_y(x,\sigma(x))|}, 
\]
so that
\[
\tau > \frac{2\pi}{c}.
\]
\item For two streamlines $(x,\sigma_1(x))$ and $(x,\sigma_2(x))$ with $\sigma_1(x) < \sigma_2(x)$ we are interested in the difference
\[
\int_0^\pi \left(\frac{1}{|\psi_y(x,\sigma_2(x))|} - \frac{1}{|\psi_y(x,\sigma_1(x))|}\right)\,dx,
\]
the positivity of which we want to prove. Let $p_{1}$ and $p_{2}$ correspond to $\sigma_{1}$ and $\sigma_{2}$, respectively. Then we need to prove exactly that
\[
\int_{0}^{\pi} (h_{p}(q,p_{2}) - h_{p}(q,p_{1}))\,dq = \int_{\Sigma} h_{pp}\,dq\,dp > 0,
\]
for $\Sigma \equiv [0,\pi] \times [p_{1},p_{2}]$. At any point either $h_{pp} \leq 0$, so that
\[
h_{pp} = -h_{q}^2 h_{pp} - h_{p}^2 h_{qq} + 2 h_{q} h_{p} h_{qp} \geq - h_{p}^2 h_{qq} + 2 h_{q} h_{p} h_{qp}, 
\]
or $h_{pp} > 0$, in which case
\[
h_{pp} \geq \half (1+h_{q}^2) h_{pp}  = \half (- h_{p}^2 h_{qq} + 2 h_{q} h_{p} h_{qp}).
\]
Here we have used the fact that $|h_{q}| \leq 1$ by assumption. It thus suffices to investigate the sign of $\int_{\Sigma}  (- h_{p}^2 h_{qq} + 2 h_{q} h_{p} h_{qp})\, dq\,dp$. By partial integration in the $q$-variable
\[
-\int_{\Sigma} h_{p}^2 h_{qq}\, dq \,dp = 2 \int_{\Sigma} h_{q} h_{p} h_{qp}\,dq \, dp,
\]
since $h_{q} = 0$ on the vertical sides of $\Sigma$. Hence, relying upon Lemma~\ref{lemma:prop}~\ref{lemma:hq} and \ref{lemma:horizontal}~\ref{lemma:hqp}, we have
\[
\int_{\Sigma} h_{pp}\,dq\,dp \geq 2 \int_{\Sigma} h_{q} h_{p} h_{qp}\,dq \, dp > 0.
\]
\item Consider the quotient
\[
\left(\int_0^\pi \frac{dx}{|\psi_y(x,\sigma_2(x))|} - \int_0^\pi \frac{dx}{|\psi_y(x,\sigma_1(x))|}\right)\bigg/\left(\sigma_2(x)-\sigma_1(x)\right), 
\]
the sign of which determines the change of $\tau$ and thus of the mean drift. The Lebesgue dominated convergence theorem can be applied to consider the limit $\sigma_2 \to \sigma_1$, being 
\begin{equation}\label{eq:sigmaderivative}
\frac{d}{d\sigma} \int_0^\pi \frac{dx}{|\psi_y(x,\sigma(x))|} = \int_0^\pi \frac{\psi_{yy}(x,\sigma(x))\,dx}{\psi_y^2(x,\sigma(x))}.
\end{equation}
Since $\psi_{yy} = -\gamma > 0$ at the bifurcation point, it follows by continuity that the expression in \eqref{eq:sigmaderivative} is positive in a neighbourhood of the trivial flow from which the nontrivial waves bifurcate.
\end{enumerate}
\end{pf*}

\section{The particle trajectories}\label{sec:trajectories} 
We are now ready to discuss what our results mean for the streamlines and particle trajectories. We shall do so with the aid of two examples. 

\subsection{Irrotational waves}
\begin{figure}
\begin{center}
\includegraphics[width=10cm,height=5cm]{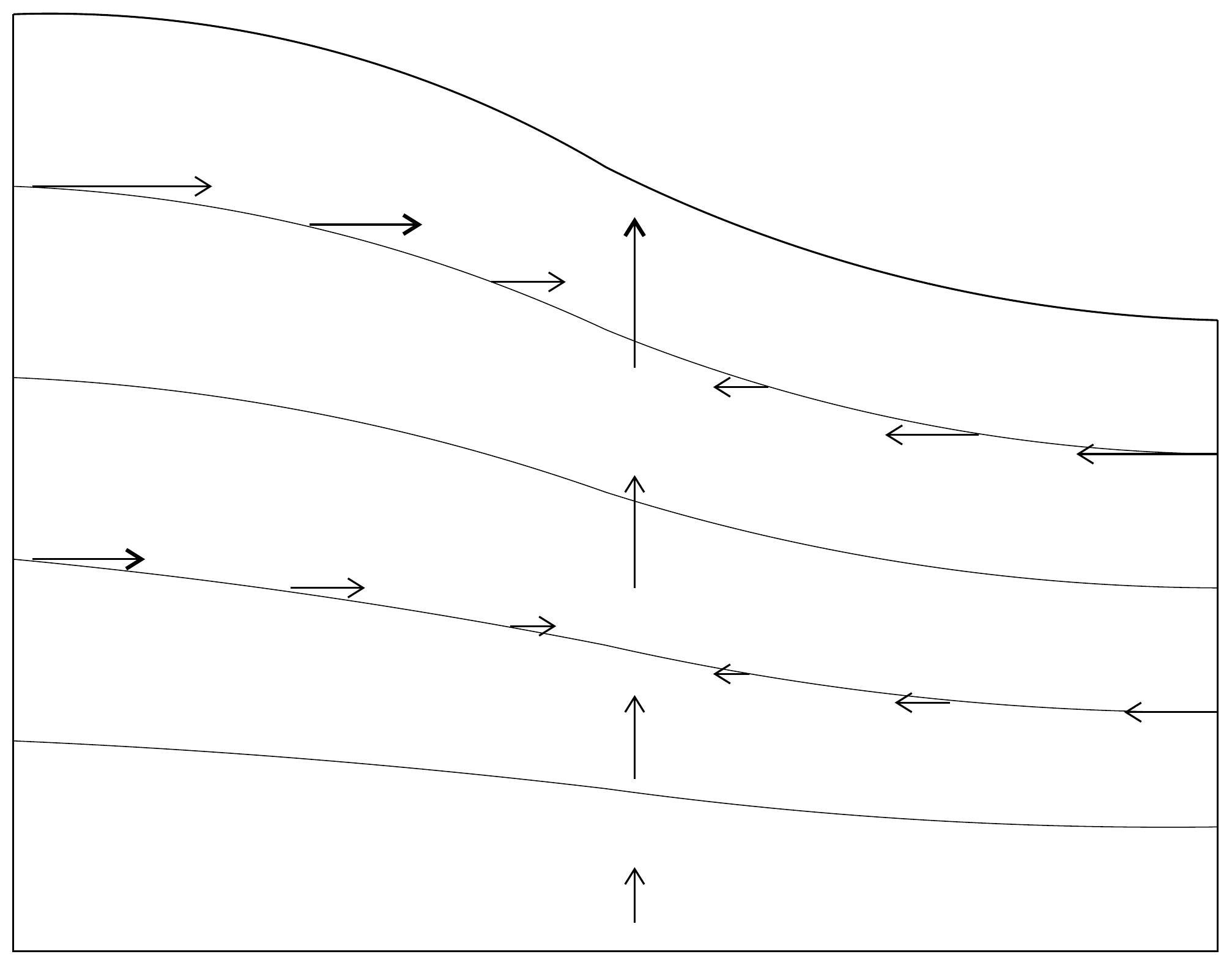}
\caption{The streamlines and the velocity field for irrotational gravity waves of small amplitude. The steepness of the streamlines and the vertical velocity is pointwise increasing from bottom to surface. The horizontal velocity is everywhere decreasing along the streamlines, and it is increasing from bottom and up beneath the crest, whilst decreasing beneath the trough. }
\label{fig:streamlines}
\end{center}
\end{figure}

Irrotational waves display extraordinary regular features (see Figure~\ref{fig:streamlines}). From bottom and up the angles between the streamlines and the horizontal plane is pointwise increasing, and for small enough waves this is true also for the vertical velocity. So is the maximal horizontal velocity, which for every streamline is attained below the crest, wherefrom it strictly decreases towards the trough. The surface is bounded below by a concave parabola, the curvature of which is determined only by the strength of gravity and the maximal horizontal velocity. 

\begin{figure}
\begin{center}
\includegraphics[width=10cm, height=6cm]{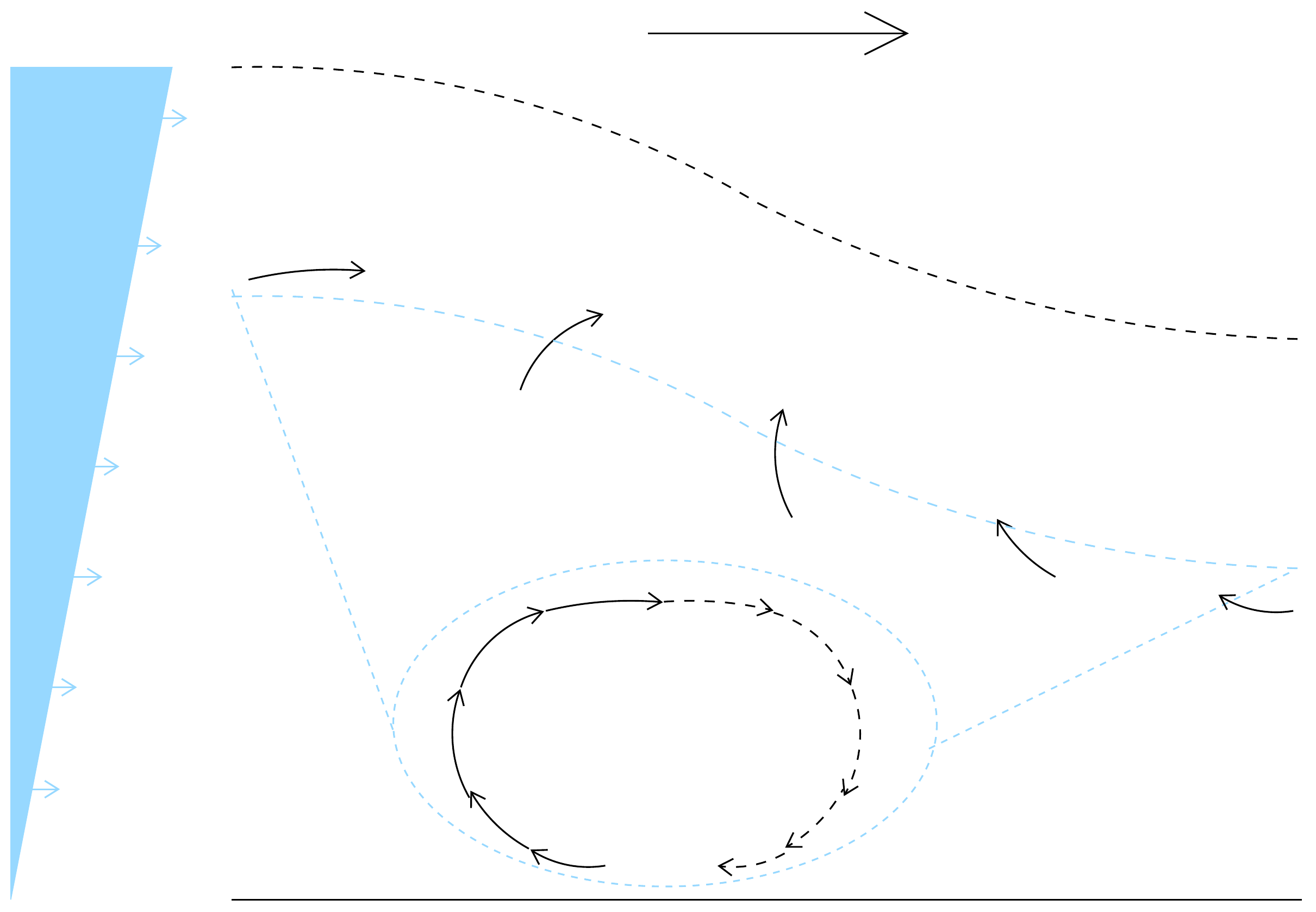}
\caption{Left: For irrotational waves and small waves of negative vorticity the forward drift is strictly increasing from bottom to surface. Center: To an observer standing still as the waves passes the particles in irrotational waves traverse non-closed elliptic orbits, corresponding to what an observer travelling along with the wave understands as a streamline. Not only the the forward drift, but also the vertical size of these orbits is increasing from bottom to surface, and the horizontal velocity of each particle is strictly increasing from the bottom of the orbit to its top. }
\label{fig:trajectories}
\end{center}
\end{figure}

In the language of particle paths, every particle traverses a non-closed circular/elliptic orbit as the wave passes above. This forward drift is strictly increasing from bottom to surface, and so is the vertical movement of the particles (this follows from the steepness of the streamlines). At the flat bed there is no forward drift; the particles move equally much back and forth. As we move up trough the fluid, each particle still moves both backward (below the trough) and forward (below the crest). As the wave propagates above, the particle moves upwards starting from the time a trough passes until the next crest passes (see Figure~\ref{fig:trajectories}). At the top of its orbit, just as the crest passes, the particle attains its maximal horizontal velocity. The movement then continues in a symmetric way, and the particle begins its descent with the horizontal speed strictly decreasing until it reaches its minimal value as the next trough passes.

\subsection{Waves of constant negative vorticity}
Some of the above features persist for waves of constant negative vorticity. The maximal steepness of the streamlines is strictly increasing from bed to surface, but we lack a proof of this property holding along any vertical line. At the surface the horizontal velocity is now non-increasing from crest to trough, and so is it at the bottom. In between we do not know. If the vorticity is positive and the wave is small, then the horizontal velocity is decreasing along any horizontal line from crest to trough, while in the half-period to the right of the crest, the vertical velocity is pointwise increasing from bottom to surface. 

The surface is bounded below by the same parabola as are irrotational waves. The fluid particles still show a mean forward drift (except at the flat bed).  The forward drift is however always strictly increasing from bed to surface in some neighbourhood of the bifurcation point (cf. \cite{MR2027299}). We have not been able to verify the circular/elliptic shape of the trajectories troughout the fluid since there is no control of the horizontal velocity in the interior of the fluid (expect below the crest). At present the possibility of particles moving constantly forward cannot be ruled out.

\section{Some comments}\label{sec:comments}
Many of the results here obtained can be extended to deep-water waves and solitary waves. However, it should be noted that there are important differences between these types of waves. Notably, the investigation \cite{MR2318158} shows that the particle trajectories within irrotational solitary waves differ in fundamental ways from those in periodic waves, and in \cite{Ehrnstrom2007d} it is proved that the class of vorticities allowed for in deep-water waves is much more restrictive than for finite depth.

The detailed information presented concerning irrotational waves is in almost all verifiable ways consistent with the behaviour of linear waves. This is expected but far from obvious, since a priori there is nothing to tell what ``closeness'' between the exact wave and its linear approximation means in terms of sharp properties. 

In the experimental study \cite{RW50} there is a picture of the particle paths within a progressive water wave. This shows very clearly two of the properties here proved mathematically: the slight forward drift and the fact that the vertical size of the orbits is decreasing with depth. Since all recognizable features in the experimental picture are consistent with our mathematical findings, it is likely that some of the more detailed properties we have established are also valid for real water waves. Hopefully, this mathematical theory can inspire the examination of such properties in laboratory settings.    
\bibliographystyle{siam}
\bibliography{thebib}

\begin{thebibliography}{10}

\bibitem{MR2257390}
{\sc A.~Constantin}, {\em The trajectories of particles in {S}tokes waves},
  Invent. Math., 166 (2006), pp.~523--535.

\bibitem{CEV06}
{\sc A.~Constantin, M.~Ehrnstr{\"o}m, and G.~Villari}, {\em Particle
  trajectories in linear deep--water waves}.
\newblock Nonlinear Anal. Real World Appl., doi:10.1016/j.nonrwa.2007.03.003.

\bibitem{CEW06}
{\sc A.~Constantin, M.~Ehrnstr{\"o}m, and E.~Wahl{\'e}n}, {\em Symmetry for
  steady gravity water waves with vorticity}, Duke Math. J.,  (2007).

\bibitem{MR2256915}
{\sc A.~Constantin and J.~Escher}, {\em Symmetry of steady periodic surface
  water waves with vorticity}, J. Fluid Mech., 498 (2004), pp.~171--181.

\bibitem{MR2318158}
\leavevmode\vrule height 2pt depth -1.6pt width 23pt, {\em Particle
  trajectories in solitary water waves}, Bull. Amer. Math. Soc. (N.S.), 44
  (2007), pp.~423--431 (electronic).

\bibitem{MR2264220}
{\sc A.~Constantin, D.~Sattinger, and W.~Strauss}, {\em Variational
  formulations for steady water waves with vorticity}, J. Fluid Mech., 548
  (2006), pp.~151--163.

\bibitem{MR2027299}
{\sc A.~Constantin and W.~Strauss}, {\em Exact steady periodic water waves with
  vorticity}, Comm. Pure Appl. Math., 57 (2004), pp.~481--527.

\bibitem{Constantin2007}
{\sc A.~Constantin and W.~Strauss}, {\em Rotational steady water waves near
  stagnation}, Phil. Trans. R. Soc. Lond. A,  (2007),
  p.~doi:10.1098/rsta.2007.2004.

\bibitem{CV06}
{\sc A.~Constantin and G.~Villari}, {\em Particle trajectories in linear water
  waves}.
\newblock J. Math. Fluid Mech., doi:10.1007/s00021-005-0214-2.

\bibitem{E06a}
{\sc M.~Ehrnstr{\"o}m}, {\em A unique continuation principle for steady
  symmetric water waves with vorticity}, J. Nonlinear Math. Phys., 13 (2006),
  pp.~484--491.

\bibitem{Ehrnstrom2007d}
{\sc M.~Ehrnstr{\"o}m}, {\em Deep-water waves with vorticity: symmetry and
  rotational behaviour}, Discrete Contin. Dyn. Syst., 19 (2007), pp.~483--491.

\bibitem{Ehrnstrom2007c}
\leavevmode\vrule height 2pt depth -1.6pt width 23pt, {\em A new formulation of
  the water wave problem for {S}tokes waves of constant vorticity}, J. Math.
  Anal. Appl.,  (2007), p.~doi:10.1016/j.jmaa.2007.07.022.

\bibitem{Ehrnstrom2007b}
{\sc M.~Ehrnstr{\"o}m and G.~Villari}, {\em Linear water waves with vorticity:
  rotational features and particle paths}.
\newblock Submitted, 2007.

\bibitem{MR1751289}
{\sc L.~E. Fraenkel}, {\em An introduction to maximum principles and symmetry
  in elliptic problems}, vol.~128 of Cambridge Tracts in Mathematics, Cambridge
  University Press, Cambridge, 2000.

\bibitem{MR2097656}
{\sc M.~D. Groves}, {\em Steady water waves}, J. Nonlinear Math. Phys., 11
  (2004), pp.~435--460.

\bibitem{23405}
{\sc D.~Henry}, {\em The trajectories of particles in deep-water {S}tokes
  waves}, International Mathematics Research Notices, 2006 (2006), pp.~Article
  ID 23405, 13 pages.
\newblock doi:10.1155/IMRN/2006/23405.

\bibitem{MR2287829}
\leavevmode\vrule height 2pt depth -1.6pt width 23pt, {\em Particle
  trajectories in linear periodic capillary and capillary-gravity deep-water
  waves}, J. Nonlinear Math. Phys., 14 (2007), pp.~1--7.

\bibitem{Hur2007}
{\sc V.~Hur}, {\em Symmetry of steady periodic water waves with vorticity},
  Philos. Trans. R. Soc. Lond. Ser. A, 365 (2007), pp.~203--2214.

\bibitem{MR1629555}
{\sc R.~S. Johnson}, {\em A modern introduction to the mathematical theory of
  water waves}, Cambridge Texts in Applied Mathematics, Cambridge University
  Press, Cambridge, 1997.

\bibitem{MR642980}
{\sc J.~Lighthill}, {\em Waves in fluids}, Cambridge University Press,
  Cambridge, 1978.

\bibitem{1109.35022}
{\sc P.~Pucci and J.~Serrin}, {\em {The strong maximum principle revisited.}},
  J. Differ. Equations, 196 (2004), pp.~1--66.

\bibitem{RW50}
{\sc F.~Ruellan and A.~Wallet}, {\em Trajectoires internes dans un clapotis
  partiel}, Houille bl., 5 (1950), pp.~483--489.

\bibitem{Stokes49}
{\sc G.~G. Stokes}, {\em On the theory of oscillatory waves}, Trans. Cambridge
  Phil. Soc., 8 (1849), pp.~441--455.

\bibitem{SCJ01}
{\sc C.~Swan, I.~Cummings, and R.~James}, {\em An experimental study of
  two-dimensional surface water waves propagating on depth-varying currents},
  J. Fluid Mech, 428 (2001), pp.~273--304.

\bibitem{MR985439}
{\sc A.~F. Teles~da Silva and D.~H. Peregrine}, {\em Steep, steady surface
  waves on water of finite depth with constant vorticity}, J. Fluid Mech., 195
  (1988), pp.~281--302.

\bibitem{MR1422004}
{\sc J.~F. Toland}, {\em {S}tokes waves}, Topol. Methods Nonlinear Anal., 7 [8]
  (1996 [1997]), pp.~1--48 [412--414].

\bibitem{Varvaruca2007}
{\sc E.~Varvaruca}, {\em On some properties of travelling water waves with
  vorticity}.
\newblock Preprint, 2007.

\bibitem{Varvaruca2007a}
\leavevmode\vrule height 2pt depth -1.6pt width 23pt, {\em On the existence of
  extreme waves and the {S}tokes conjecture with vorticity}.
\newblock Preprint available on arXiv:0707.2224, 2007.

\bibitem{MR2130838}
{\sc E.~Wahl{\'e}n}, {\em A note on steady gravity waves with vorticity}, Int.
  Math. Res. Not.,  (2005), pp.~389--396.

\bibitem{wahlen:921}
\leavevmode\vrule height 2pt depth -1.6pt width 23pt, {\em Steady periodic
  capillary-gravity waves with vorticity}, SIAM Journal on Mathematical
  Analysis, 38 (2006), pp.~921--943.

\end{thebibliography}

\end{document}